\documentclass[preprint,12pt]{elsarticle}

\usepackage{amssymb}

\usepackage{braket}
\usepackage{amsmath}
\usepackage{amsfonts}

\journal{Advances in Quantum Chemistry}

\begin{document}

\begin{frontmatter}



\title{Multicenter integrals involving complex Gaussian type functions}


\author{Abdallah Ammar, Arnaud Leclerc and Lorenzo Ugo Ancarani}

\address{Universit\'e de Lorraine, CNRS, Laboratoire de Physique et Chimie Th\'eoriques,
	57000 Metz, \underline{France}}

\begin{abstract}
Multicentric integrals that involve a continuum state cannot be evaluated with the usual quantum chemistry tools and require a special treatment.
We consider an initial molecular bound state described by  multicenter spherical or cartesian Gaussian functions.
An electron ejected through an ionization process will be described by an oscillating continuum wavefunction that enters the matrix element necessary for cross section calculations.
Within a monocentric approach, we have recently shown how such integrals can be evaluated analytically
by using a representation of the continuum state by a set of complex Gaussian functions.
In this work we tackle the multicentric situation.
The method, developed in either spherical or cartesian coordinates, and validated by numerical tests, makes use of existing mathematical tools extended to the complex plane.
\end{abstract}


\end{frontmatter}


\section{Introduction}

There exists a vast literature dedicated to the evaluation of integrals necessary in quantum chemistry and in molecular physics.
According to the type of functions involved (like Gaussian-type or Slater-type, correlated or not correlated, and in diverse coordinate systems)
specific strategies may be envisaged.
Within the realm of bound states, in particular, Gaussians play a particular role since they are commonly used by a large number of chemists
and physicists alike \cite{Jensen2013}.
The main reason for this widespread use is that Gaussians possess very nice properties, including the famous
product theorem \cite{shavitt1963}; they greatly facilitate the evaluation of multicenter integrals that appear in molecular studies.

When continuum states (or even highly excited states) are involved, as in scattering problems, integrals are more difficult to
evaluate because of oscillating integrands spanning generally a more extended radial range. Besides, Gaussian-type functions are
not a natural basis choice for continuum states. Indeed, radial functions oscillating  up to infinity are difficult to represent by fast decaying Gaussians.
This option has been explored only in  few studies
\cite{kaufmann1989universal, nestmann1990optimized, faure2002gtobas},
since least square real Gaussian representations turn out to be relatively inefficient
 in reproducing the oscillations at large radial distance
and high energy (typically $r \sim 15-20$  a.u. and $E \sim 1.5-2$ a.u. for a regular Coulomb function, as an example).
Alternatively, complex Gaussians, \textit{i.e.} Gaussian functions with complex exponents,
could be more appropriate for continuum calculations because they intrinsically oscillate.
Since McCurdy and Rescigno \cite{mccurdy1978extension} proposed their use
as an alternative to the complex-coordinate technique, complex Gaussians have been
applied to study ionization cross sections
\cite{rescigno1985locally, yu1985molecular, yabushita1987complex, morita2008calculations, morita2008photoionization, morita2008photoionization2, matsuzaki2017calculation}
and molecular resonances \cite{honigmann2010use, white2015complex},
and even for non-Born-Oppenheimer ground and
excited states calculations (see \cite{bubin2020computer} and references therein).
In a recent study \cite{Ammar2020,AmmarPhD}, we have developed an efficient numerical approach to represent
a set of oscillating functions over a finite range by optimized complex Gaussians.
We have also performed a detailed comparison between using complex and real Gaussians, and we can summarize our findings as follows:
complex Gaussians are better suited for oscillating functions (in particular for larger energies),
allow one to cover larger radial domains with greater accuracy and avoid unphysical behaviors outside the fitting box.
Returning to the goal of evaluating efficiently integrals involved in scattering processes,
we have shown \cite{Ammar2020,AmmarPhD} that
a complex Gaussian representation of continuum states allows for an analytical approach, as long as one stays within a monocentric framework.



The idea of the present work is to show how to extend the method when matrix elements include multicentric molecular integrals over continuum states. If the bound state is described in terms of real Gaussians, our continuum state representation by optimized complex Gaussians leads to an all-Gaussian approach that can take advantage of the Gaussian properties, similar to what is done in quantum chemistry.
In the one active electron approach
we wish to compute efficiently matrix elements like
\begin{equation}
\mathcal{T}_{\mathbf{k_e}} =
\braket{\psi_{\mathbf{k_e}}\left(\mathbf{r}-\mathbf{O}\right)|\widehat{O}|
	\Phi_{i}\left(\mathbf{r}-\mathbf{R}\right)},
\label{eq:transition_elementGENERAL}
\end{equation}
where $\Phi_{i}$ is an initial bound state centered around a position $\mathbf{R}$, $\psi_{\mathbf{k_e}}$ is a continuum state (with a wavevector $\mathbf{k_e}$) centered around a position $\mathbf{O}$,
and $\widehat{O}$ is an operator.
To illustrate the usefulness of representing continuum with complex Gaussians,  we have started with a simpler problem, in which both states are centered on the origin of the coordinate system \cite{Ammar2020,AmmarPhD}.
Also, to compare with a solid and exact benchmark, we first considered the cross section calculation of the photoionization of atomic hydrogen for which all wavefunctions involved are known analytically ($\widehat{O}$ is just the dipole operator). Through a complex Gaussian representation of the radial part of the continuum, and a real Gaussian expansion of a given exact initial state $nl$, the matrix element becomes fully analytical. Actually it turns out that, although mathematically different, the matrix element calculation is also analytical if the initial state is described by Slater-type orbitals. Next, we have considered the photoionization of molecules, whose initial and
final states are described in a one-centre approach \cite{AmmarPhD,Ammar2021}.
The strategy works as well when the operator is replaced
by a plane wave ($e^{\imath \mathbf{q} \cdot \mathbf{r}}$) that appears, within the
first Born approximation, when evaluating cross sections for ionization by a charged
particle impact ($\mathbf{q}$ is the momentum transferred to the target).
In all cases,  the numerically demanding part is the non-linear optimization
of complex Gaussians to represent the continuum. The matrix elements, expressed as
multiple limited sums involving special functions, are easily computed.
The successful numerical comparison illustrated that the proposed
all-Gaussian approach works efficiently for ionization processes of one-center targets.

  In the present paper, we wish to show that we can go a step further towards a multicentric case, still retaining the analytical character of the approach.
 Following the quantum chemistry strategy used for bound states we make use of Gaussians properties. When applying the product theorem with complex Gaussians, though, one major issue is that we are faced with complex centers.
Thus, the study of such matrix elements is not a mere extension
of the monocentric case  but involves some more advanced mathematical
tools that we present hereafter.

In the next section we provide the proposed formulation for either spherical or
cartesian Gaussian functions.
A numerical test case is then presented in Section 3, and our findings are summarized in Section 4.



\section{Analytical evaluation of transition matrix elements}

Consider the photoionization of a molecule. Initially in the electronic state $\Phi_{i}$,
it is ionized by a photon of energy $\omega$ and a polarization vector $\epsilon$, say, chosen along of z-axis ($\epsilon = \hat{z}$).
We suppose that the vibrational and rotational structure of molecule is
not affected in the process. Let the ejected electron have an energy $k_e^2/2$ and
a wavevector $\mathbf{k_e}$. Its continuum state
$\psi_{\mathbf{k_e}}$ is supposed, for simplicity, to be
centered on the origin $\mathbf{O}=\mathbf{0}$.
The initial molecular state will have several centers. However, for our purposes, it will be sufficient
to take
a single function $\Phi_{i}\left(\mathbf{r}-\mathbf{R}\right)$ centered on a vector
$\mathbf{R} \equiv  \left( X,Y,Z \right)$.

Within the dipole approximation, the transition moment
is given by
\begin{equation}
\mathcal{T}_{\mathbf{k_e}}^{(g)} =
\braket{\psi_{\mathbf{k_e}}\left(\mathbf{r}\right)|\widehat{O}^{(g)}|
	\Phi_{i}\left(\mathbf{r}-\mathbf{R}\right)},
\label{eq:transition_element}
\end{equation}
where the dipole operator in length $(L)$ or velocity $(V)$ gauge reads
\begin{equation}
\begin{aligned}
\widehat{O}^{(L)} &= \hat{\epsilon} \cdot \mathbf{r} = z
=  \sqrt{\frac{4 \pi}{3}} r Y_1^0\left(\hat{r}\right), \\
\widehat{O}^{(V)} &= -\frac{\imath}{w} \hat{\epsilon} \cdot \mathbf{p} =  -\frac{\imath}{\omega} p_z,
\end{aligned}
\label{eq:Operat_g}
\end{equation}
where $Y_l^m$ is the usual spherical harmonic.
If exact wavefunctions are used, the two gauges are equivalent, and results of the calculations coincide.
In this paper, we shall consider only the length gauge formulation.

We expand the continuum wavefunction in partial waves
\begin{equation}
\psi_{\mathbf{k_e}}(\mathbf{r})
= \sqrt{\frac{2}{\pi}}
\sum_{l=0}^{\infty} \sum_{m=-l}^{l} (\imath)^l
e^{-\imath \sigma_l} \frac{u_{l,k_e}(r)}{k_e r}
Y_l^m(\hat{r}) Y_l^{m*}\left(\widehat{k_e}\right) ,
\label{eq:continuum_state}
\end{equation}
where the radial
functions $u_{l,k_e}(r)$ are the solutions of the radial Schr\"odinger
equation, and $\sigma_l$ are the corresponding phase-shifts.
As suggested in~\cite{Ammar2020}, these radial functions can be represented by a set of
$N$ optimized complex Gaussians,
\begin{equation}
\left(u_{l,k_e}(r)\right)^* = r^{l+1} \sum_{s=1}^N
\left[ c_s \right]_{l,k_e} e^{-\left[\alpha_s\right]_l r^2} .
\label{eq:optimized_cG}
\end{equation}
The non-linear exponents $\left[\alpha_s\right]_l$ and the linear coefficients $\left[ c_s \right]_{l,k_e}$ are here supposed to be already known.

For the initial state $\Phi_{i}\left(\mathbf{r}-\mathbf{R}\right)$, we consider here two different Gaussian-type
functions: $(i)$ the spherical Gaussian-type functions (SGTFs)
in section~\ref{sec:spheric_Gauss}; and
$(ii)$ the cartesian Gaussian-type functions (CGTFs)
in section~\ref{sec:cartes_Gauss}.
In both cases, we will show that the matrix element~(\ref{eq:transition_element})
can be evaluated analytically.
We also emphasize here that if $\mathbf{R}=\mathbf{0}$, that is to say in
the monocentric case, the formulation developed hereafter reduces to that
presented in \cite{Ammar2020, AmmarPhD}.


\subsection{Spherical Gaussian-type functions}\label{sec:spheric_Gauss}

We suppose here that the initial state is a
SGTF
\begin{equation}
\Phi_{n_i l_i m_i}(\mathbf{r}) =
\mathbb{Y}_{l_i m_i}^{n_i}(\mathbf{r})
e^{-\beta_i r^2},
\label{eq:SGTF}
\end{equation}
where
\begin{equation}
\mathbb{Y}_{l_i m_i}^{n_i}(\mathbf{r}) =
\left( \mathbf{r} \cdot \mathbf{r} \right)^{n_i}
\mathcal{Y}_{l_i}^{m_i}(\mathbf{r}),
\label{eq:Harm_Polyn}
\end{equation}
is the general harmonic polynomial and
$\mathcal{Y}_{l}^{m}$ is the solid harmonic.

For real argument $\mathbf{r} \in \mathbb{R}^3$, the latter is
given in terms of the usual spherical harmonic
\begin{equation}
\mathcal{Y}_{l}^{m}(\mathbf{r}) = r^l Y_{l}^{m}(\hat{r}) ,
\end{equation}
while for complex arguments $\mathbf{r} \equiv (x,y,z) \in \mathbb{C}^3$, it
becomes~\cite{biedenharn_1981}
\begin{equation}
\mathcal{Y}_l^{m}\left( \mathbf{r} \right) =
\sqrt{\frac{2l+1}{4 \pi} (l+m)! (l-m)!}
\sum_k \frac{(-x-\imath y)^{k+m} (x-\imath y)^{k} (z)^{l-2k-m}}
{2^{2k+m}(k+m)! \, k! \, (l-m-2k)!}.
\end{equation}
Note that the integer number $n_i$ is not a quantum number; it allows one to consider different types of Gaussian functions.
The very commonly employed s-type Gaussian corresponds to $n_i=l_i=0$.

Molecular integrals involving SGTF
with real arguments were
considered by Harris~\cite{harris1963gaussian} and
Krauss~\cite{krauss1964gaussian}; they can be performed
analytically by repeatedly rotating the spherical harmonics.
Fieck~\cite{fieck1979racah} evaluated multicenter integrals
by using the Talmi transformation and the
Moshinsky-Smirnov coefficients of nuclear shell theory.
Kaufmann and Baumeister~\cite{kaufmann1989single} derived a
single center expansion for an arbitrary Gaussian  and used
the modified Rayleigh expansion.
Chiu and Moharerrzadeh~\cite{chiu1994translational} applied a similar
strategy in which they used a
translational and rotational expansion of the
Gaussian from one center to another.
They then treated the two-electron integrals
by using the Fourier transformation convolution
theorem~\cite{moharerrzadeh1996multicenter}.
Kuang and Lin~\cite{kuang1997molecularI} derived analytical
results for multicenter integrals over SGTFs by using
the Fourier transform and the addition theorem of
harmonic polynomials. They also apply the same approach to treat
molecular integrals involving also a plane wave~\cite{kuang1997molecularII}.
For this purpose, they generalized the
addition theorem of harmonic polynomials for complex arguments.
In this section, we will use this generalization to obtain an analytical
expression for the transition element~(\ref{eq:transition_element}).
We will first consider the special case $n_i=0$ which is very common
in quantum chemistry and allows one to master the necessary mathematical steps.
Then we will move on to the general case $n_i\geq 0$.


\subsubsection{Special case: $n_i=0$}

We consider here the transition matrix element~(\ref{eq:transition_element}) (in the length gauge)
\begin{equation}
\label{sphericalME}
\begin{aligned}
\mathcal{T}^{(L)}_{\mathbf{k_e}}
&=
\braket{\psi_{\mathbf{k_e}}\left(\mathbf{r}\right)
	| \sqrt{\frac{4 \pi}{3}} r Y_1^0\left(\hat{r}\right)
	|\Phi_{n_i=0 \, l_i m_i}\left(\mathbf{r}-\mathbf{R}\right)} \\
&=\frac{2\sqrt{2}}{k_e \sqrt{3}}
\sum_{l,m} (-\imath)^l e^{\imath \sigma_l} Y_l^m \left( \widehat{k_e} \right)
\mathcal{J}_{l,m},
\end{aligned}
\end{equation}
with
\begin{equation}
\begin{aligned}
\mathcal{J}_{l,m}
&= \int d \mathbf{r}
\left(u_{l,k_e}(r)\right)^*  Y_l^{m*}(\hat{r}) \, Y_1^0(\hat{r}) \,
e^{-\beta_i \left( \mathbf{r} - \mathbf{R} \right)^2}
\mathcal{Y}_{l_i}^{m_i} \left( \mathbf{r} - \mathbf{R}\right) .
\label{eq:integ_J_def}
\end{aligned}
\end{equation}
To evaluate this three-dimensional integral, we make use of two mathematical ingredients.
First, the addition theorem for solid harmonics~\cite{steinborn1973rotation}
\begin{equation}
\mathcal{Y}_{l_i}^{m_i} (\mathbf{r} - \mathbf{R})
= 4 \pi \sum_{l'=0}^{l_i} \sum_{m'=m_{\min}'}^{m_{\max}'}
G(l_i \, m_i | l' \, m')
\mathcal{Y}_{l'}^{m'} (\mathbf{r} ) \mathcal{Y}_{l_i-l'}^{m_i-m'} (-\mathbf{R} ),
\label{eq:SH_add}
\end{equation}
with $m_{\min}'= \max(-l',m_i-l_i+l')$,
$m_{\max}'= \min(l',m_i+l_i-l')$, and
\begin{equation}
G(l_i \, m_i | l' \, m')  =
\frac{(2l_i+1)!!}{(2l'+1)!!\left[ 2(l_i-l')+1 \right]!!}
\braket{ l_i \, m_i | l' \, m' | l_i-l' \, , m_i-m' }
\end{equation}
where $\braket{ l_2 \, m_2 | L \, M | l_1 \, m_1 }$ denotes the
Gaunt coefficient
\begin{equation}
\braket{ l_2 \, m_2 | L \, M | l_1 \, m_1 }
= \int d \hat{r} Y_{l_2}^{m_2 *}\left(\hat{r}\right)
Y_{L}^{M}\left(\hat{r}\right) Y_{l_1}^{m_1}\left(\hat{r}\right).
\end{equation}
Second, we need the expansion formula \cite{kuang1997molecularII}
\begin{equation}
e^{2 \beta_i \mathbf{r} \cdot \mathbf{R}} =
4 \pi \sum_{\lambda=0}^{\infty} \sum_{\mu=-\lambda}^{\lambda}
(-1)^{\mu}
\mathcal{Y}_{\lambda}^{\mu}\left(\mathbf{r}\right)
\mathcal{Y}_{\lambda}^{-\mu}\left(2 \beta_i \mathbf{R}\right)
\sum_{k=0}^{\infty}
\frac{
	\left( \mathbf{r} \cdot \mathbf{r}  \right)^k
	\left[ 2 \beta_i \mathbf{R} \cdot 2 \beta_i \mathbf{R}  \right]^k}
{(2k)!!(2k+2\lambda+1)!!} .
\label{eq:Gen_Rayl_expansion}
\end{equation}
Applying the homogeneous property
$\mathcal{Y}_l^m\left( \gamma \mathbf{r} \right) = \gamma^l \mathcal{Y}_l^m\left(\mathbf{r} \right)$,
we can write \newline
$\mathcal{Y}_{\lambda}^{-\mu}\left(2 \beta_i \mathbf{R}\right)
 = 2^{\lambda} \mathcal{Y}_{\lambda}^{-\mu}\left(\beta_i \mathbf{R}\right)$.

Inserting formulae (\ref{eq:SH_add}) and (\ref{eq:Gen_Rayl_expansion}), and replacing the radial function of the continuum
by its Gaussian representation (\ref{eq:optimized_cG}), the integral (\ref{eq:integ_J_def}) becomes
\begin{equation}
\begin{aligned}
\mathcal{J}_{l,m}
&= \left(4 \pi\right)^2 e^{-\beta_i R^2} \sum_{s=1}^N \left[ c_s \right]_{l,k_e}
\sum_{l'=0}^{l_i} \sum_{m'=m_{\min}'}^{m_{\max}'}
G(l_i \, m_i | l' \, m')  \mathcal{Y}_{l_i-l'}^{m_i-m'} (-\mathbf{R} ) \\
& \quad \times
\sum_{\lambda=0}^{\infty} \sum_{\mu=-\lambda}^{\lambda}
(-1)^{\mu} 2^{\lambda}
\mathcal{Y}_{\lambda}^{-\mu} \left(\beta_i \mathbf{R}\right)
\int d \hat{r} \, Y_l^{m*}(\hat{r}) \, Y_1^0(\hat{r}) \,
{Y}_{l'}^{m'} (\hat{r} ) {Y}_{\lambda}^{\mu}\left(\hat{r}\right) \\
& \quad \times \sum_{k=0}^{\infty}
\frac{\left[ 2 \beta_i \mathbf{R} \cdot 2 \beta_i \mathbf{R}  \right]^k}
{(2k)!!(2k+2\lambda+1)!!}
\int_0^{\infty} d{r} \, r^{l+l'+\lambda+2k+3}
e^{-\left( \left[ \alpha_s \right]_{l} + \beta_i  \right) r^2}.
\end{aligned}
\end{equation}
The angular integral over four spherical
harmonics can be seen to give
\begin{equation}
\sum_{d=|l'-1|}^{l'+1}
\braket{ l \, m | \lambda \, \mu | d \, , m-\mu }
\braket{ l' \, m' | 1 \, 0 |d \, m' }
\delta_{m',m-\mu},
\end{equation}
with the conditions $|l-d|\leq \lambda \leq l+d$, $l'+1+d=\text{even integer}$ ($d \neq l'$)
and  $l+\lambda+d=\text{even integer}$. The  radial integral is quite straightforward, and the summation over $k$
can be carried out, and yields a confluent hypergeometric
function $_1F_1$ \cite{abramowitz_stegun}, namely
\begin{equation}
\begin{aligned}
&\sum_{k=0}^{\infty}
\frac{\left[ 2 \beta_i \mathbf{R} \cdot 2 \beta_i \mathbf{R}  \right]^k}
{(2k)!!(2k+2\lambda+1)!!}
\int_0^{\infty} d{r} \, r^{l+l'+\lambda+2k+3}
e^{-\left( \left[ \alpha_s \right]_{l} + \beta_i  \right) r^2} \\
=&\sum_{k=0}^{\infty}
\frac{\left[ 2 \beta_i \mathbf{R} \cdot 2 \beta_i \mathbf{R}  \right]^k}
{\frac{2^{2k+\lambda+1}}{\sqrt{\pi}}
	\Gamma \left( k+1 \right)
	\Gamma \left( k+\lambda+\frac{3}{2} \right)}
\frac{ \Gamma \left( k + \frac{l+l'+\lambda+4}{2} \right) }
{2 \left( \left[ \alpha_s \right]_{l} + \beta_i  \right)^{k + \frac{l+l'+\lambda+4}{2}}} \\
=& \frac{\sqrt{\pi} \, \Gamma \left( \frac{l+l'+\lambda+4}{2} \right)}
{2^{\lambda+2} \Gamma \left( \lambda+\frac{3}{2} \right)
	\left( \left[ \alpha_s \right]_{l} + \beta_i  \right)^{\frac{l+l'+\lambda+4}{2}}} \,
_1F_1\left( \frac{l+l'+\lambda+4}{2},\lambda+\frac{3}{2} ;
\frac{\beta_i^2 R^2}{ \left[ \alpha_s \right]_{l} + \beta_i }\right).
\end{aligned}
\end{equation}

Collecting the results of the two integrations gives us the following final expression
for the integral $\mathcal{J}_{l,m} $
\begin{equation}
\begin{aligned}
\mathcal{J}_{l,m}
&=  4 \pi^2 \sqrt{\pi} (-1)^{m} \,(2l_i+1)!! \, e^{-\beta_i R^2}
\sum_{l'=0}^{l_i}
\frac{1}{(2l'+1)!!\left[ 2(l_i-l')+1 \right]!!}  \\
& \quad \times
\sum_{m'=m_{\min}'}^{m_{\max}'}  (-1)^{m'}
\braket{ l_i \, m_i | l' \, m' | l_i-l' \, , m_i-m' }
\mathcal{Y}_{l_i-l'}^{m_i-m'} (-\mathbf{R} ) \label{integralnequal0} \\
& \quad \times
\sum_{d=|l'-1|}^{l'+1} \braket{ l' \, m' | 1 \, 0 |d \, m' }
\sum_{\lambda=|l-d|}^{l+d} \braket{ l \, m | \lambda \, , m-m' | d \, , m' }  \\
& \quad \times
\frac{\Gamma \left( \frac{l+l'+\lambda+4}{2} \right)}{\Gamma \left( \lambda+\frac{3}{2} \right)}
\mathcal{G}_{l \, l' \, \lambda}(k_e)
\mathcal{Y}_{\lambda}^{m'-m}\left(\beta_i \mathbf{R}\right),
\end{aligned}
\end{equation}
with
\begin{equation}
\mathcal{G}_{l \, l' \, \lambda}(k_e) =
\sum_{s=1}^N
\frac{\left[ c_s \right]_{l,k_e}}{\left( \left[ \alpha_s \right]_{l} + \beta_i  \right)^{\frac{l+l'+\lambda+4}{2}}}\,
_1F_1\left( \frac{l+l'+\lambda+4}{2},\lambda+\frac{3}{2} ;
\frac{\beta_i^2 R^2}{ \left[ \alpha_s \right]_{l} + \beta_i }\right).
\label{eq:G_def}
\end{equation}
The elements (\ref{eq:G_def}) can be easily evaluated from the Gaussian exponents associated with the initial $\beta_i$ and continuum $\left[ \alpha_s \right]_{l}$ states.

Note that in the development the key stages~(\ref{eq:SH_add}) and~(\ref{eq:Gen_Rayl_expansion}) have been applied to real arguments.
However, it is important to realize that they are applicable also for complex
arguments~\cite{kuang1997molecularII}, a situation that appears if the continuum wavefunction
is not centered on the origin ($\mathbf{O} \neq \mathbf{0}$); complex arguments would appear also when the dipole operator $\widehat{O}^{(L)} $ is replaced by a plane wave,
for example when studying the ionization by charged particle
impact within the first Born approximation.


\subsubsection{General case: $n_i \geq 0$}

The integral~(\ref{eq:integ_J_def}) reads now
\begin{equation}
\mathcal{J}_{l,m}
= \int d \mathbf{r}
\left(u_{l,k_e}(r)\right)^*  Y_l^{m*}(\hat{r}) \, Y_1^0(\hat{r}) \,
e^{-\beta_i \left( \mathbf{r} - \mathbf{R} \right)^2}
\mathbb{Y}_{l_i \, m_i}^{n_i} \left( \mathbf{r} - \mathbf{R}\right).
\label{eq:integ_J_def_n}
\end{equation}
In this case, we use the addition theorem of the general harmonic
polynomial, the generalization of~(\ref{eq:SH_add}), namely~\cite{niukkanen1980transformation,niukkanen1983sigma}
\begin{equation}
\begin{aligned}
\mathbb{Y}_{l_i m_i}^{n_i} (\mathbf{r} - \mathbf{R})
&= 4 \pi
\sum_{n_1=0}^{n_i-\Delta}
\sum_{l_1=0}^{l_i+2n_i}
\sum_{l_2=|l_i-l_1|}^{(l_2)_{\max}}
\sum_{m_1=(m_1)_{\min}}^{(m_1)_{\max}}
\mathcal{K}(\mathbf{n_i}|\mathbf{n_1}|\mathbf{n_2}) \\
& \quad  \quad \quad \times
\mathbb{Y}_{l_1 m_1}^{n_1}(\mathbf{r})
\mathbb{Y}_{l_2 m_2}^{n_2} (\mathbf{-R}),
\end{aligned}
\label{eq:PolHar_add}
\end{equation}
with the notation $\mathbf{n} \equiv (n,l,m)$  and
\begin{equation}
\mathcal{K}(\mathbf{n_i}|\mathbf{n_1}|\mathbf{n_2}) =
\frac{(2n_i)!!}{(2n_1)!! (2n_2)!!}
\frac{[2(n_i+l_i)+1]!! \braket{l_i \, m_i | l_1 \, m_1 | l_2 \, m_2 } }
{[2(n_1+l_1)+1]!! [2(n_2+l_2)+1]!!},
\end{equation}
where $\Delta = (l_1+l_2-l_i)/2$,
$n_2=n_i-n_1-\Delta$,
$(l_2)_{\max} =  \min(l_1+l_i,l_i+2n_i-l_1)$,
$\max(-l_1,m_i-l_2) \le m_1 \le \min(l_1,m_i+l_2)$,
$m_2 = m_i - m_1$; the summations are restricted to  $l_1 + l_2 + l_i$ being an even integer.

Therefore, by
substituting~(\ref{eq:optimized_cG}),~(\ref{eq:Gen_Rayl_expansion})
and~(\ref{eq:PolHar_add}) in~(\ref{eq:integ_J_def_n}),
we obtain
\begin{equation}
\begin{aligned}
\mathcal{J}_{l,m}
&= 4 \pi^2 \sqrt{\pi} \,
(-1)^m (2n_i)!! [2(n_i+l_i)+1]!!\,
e^{-\beta_i R^2} \\
& \quad \times
\sum_{n_1=0}^{n_i-\Delta} \frac{1}{(2n_1)!! (2n_2)!!}
\sum_{l_1=0}^{l_i+2n_i} \frac{1}{[2(n_1+l_1)+1]!!} \\
& \quad \times
\sum_{l_2=|l_i-l_1|}^{(l_2)_{\max}}
\frac{1}{[2(n_2+l_2)+1]!!}
\mathbb{Y}_{l_2 m_2}^{n_2} (\mathbf{-R}) \\
& \quad \times
\sum_{m_1=(m_1)_{\min}}^{(m_1)_{\max}} (-1)^{m_1}
\braket{l_i \, m_i | l_1 \, m_1 | l_2 \, m_2 }  \\
& \quad \times
\sum_{d=|l_1-1|}^{l_1+1}
\braket{ l_1 \, m_1 | 1 \, 0 |d \, m_1 }
\sum_{\lambda=|l-d|}^{l+d}
\braket{ l \, m | \lambda \, , m-m_1 | d \, m_1 }  \\
& \quad \times
\frac{\Gamma \left( \frac{l+l_1+2n_1+\lambda+4}{2} \right)}
{\Gamma \left( \lambda + \frac{3}{2} \right)}
\mathcal{G}_{l \,  l_1+2n_1 \,  \lambda}(k_e) \,
\mathcal{Y}_{\lambda}^{m_1-m}\left(\beta_i \mathbf{R}\right),
\label{ni ne 0}
\end{aligned}
\end{equation}
where angular and radial integrals are performed
similarly to the $n_i=0$ case,
and $\mathcal{G}_{l \,  l_1+2n_1 \,  \lambda}(k_e)$
is given by~(\ref{eq:G_def}) with $l_1+2n_1$ instead of $l'$.

Note that since the addition theorem (\ref{eq:PolHar_add}) is valid also for complex arguments~\cite{kuang1997molecularII},
the formulation proposed by (\ref{ni ne 0}) still applies for example if the continuum is centered on another position.


\subsection{Cartesian Gaussian-type functions}\label{sec:cartes_Gauss}

We suppose now that the initial state is a CGTF of the form
\begin{equation}
\Phi_{n_x \, n_y \, n_z}(\mathbf{r}) =
x^{n_x} y^{n_y} x^{n_z} e^{-\beta_i r^2}.
\label{eq:CGTF}
\end{equation}
CGTFs have been employed first by Boys~\cite{boys1950electronic} in
the calculation of molecular integrals.
He showed~\cite{boys1950electronic}  that multicenter integrals
over s-CGTF ($n_x+n_y+n_z=0$) can be evaluated in a
closed form, and from this result integrals involving CGTFs with
higher angular momentum can be obtained by a differentiation procedure.
Taketa \emph{et al}~\cite{taketa1966gaussian} derived an
analytical expression for molecular integrals over general CGTFs
by exploiting the Gaussians product theorem, whereby
the product of two s-CGTF centered over two different centers
gives a new s-CGTF centered over a new position. An expansion
of the linear terms over this center then yields analytical
expressions for molecular integrals.
Here we will show that we can use such a strategy to evaluate in closed form the
transition matrix elements involving complex Gaussians.

The transition matrix element~(\ref{eq:transition_element}), again in the length gauge,
is given by
\begin{equation}
\label{cartesianME}
\begin{aligned}
\mathcal{T}^{(L)}_{\mathbf{k_e}}
&=
\braket{\psi_{\mathbf{k_e}}\left(\mathbf{r}\right)
	|z|\Phi_{n_x \, n_y \, n_z}\left(\mathbf{r}-\mathbf{R}\right)} \\
&=\frac{1}{k_e} \sqrt{\frac{2}{\pi}}
\sum_{l,m} (-\imath)^l e^{\imath \sigma_l} Y_l^m \left( \widehat{k_e} \right)
\mathcal{I}_{l,m},
\end{aligned}
\end{equation}
with
\begin{equation}
\mathcal{I}_{l,m} =
\int d \mathbf{r}
\frac{\left(u_{l,k_e}(r)\right)^*}{r}  Y_l^{m*}(\hat{r})
\, z \,
\left(x-X\right)^{n_x} \left(y-Y\right)^{n_y} \left(z-Z\right)^{n_z}
e^{-\beta_i \left( \mathbf{r} - \mathbf{R} \right)^2}.
\label{eq:integ_I_def}
\end{equation}
To deal with this three-dimensional integral we need the following mathematical ingredients.
First we recall the Gaussians product theorem~\cite{boys1950electronic}
\begin{equation}
\label{product}
e^{-\left[\alpha_s\right]_l r^2}
e^{-\beta_i \left( \mathbf{r} - \mathbf{R} \right)^2}  =
e^{-\frac{\beta_i \left[\alpha_s\right]_l}
	{\beta_i + \left[\alpha_s\right]_l} R^2} \,
e^{- \left[\gamma_s\right]_l
	\left( \mathbf{r} - \left[ \mathbf{{C}_s} \right]_l \right)^2 },
\end{equation}
with
\begin{equation}
\left\lbrace
\begin{aligned}
&\left[\gamma_s\right]_l
= \beta_i + \left[\alpha_s\right]_l \in \mathbb{C}
\quad \text{with } \Re\left(\left[\gamma_s\right]_l\right) > 0, \\
& \left[ \mathbf{{C}_s} \right]_l \equiv
\left(
\left[ x_{C_s} \right]_l,\left[ y_{C_s} \right]_l,\left[ z_{C_s} \right]_l
\right)
= \frac{\beta_i}{\beta_i+\left[\alpha_s\right]_l} \mathbf{R} \in \mathbb{C}^3.
\end{aligned}
\right.
\end{equation}
The new centers $\left[ \mathbf{{C}_s} \right]_l$ are mere mathematical artifacts and do not exist physically.
Moreover, unlike the usual quantum chemistry framework, in our case the use of complex exponents $\left[\alpha_s\right]_l$ sets these artificial centers in the complex plane. We will
show hereafter that the imaginary part of this center
has no consequences on the integral.
The second ingredient is the identity~\cite{schlegel1995transformation}
\begin{equation}
\label{cartesianHarmo}
\begin{aligned}
r^l Y_l^{m*}(\hat{r})
& =
\sqrt{\frac{2l+1}{4 \pi} \, \frac{(l-|m|)!}{(l+|m|)!}}  \, \,
\sum_{k=0}^{[\frac{l-|m|}{2}]}   \mathcal{C}_1(l,|m|,k) \,
\sum_{h=0}^{k} \binom{k}{h} \,
\sum_{j=0}^{k-h} \binom{k-h}{j} \\
& \quad \times
\sum_{p=0}^{|m|} \, \binom{|m|}{p} \, \left(\mathcal{C}_2(m,p)\right)^* \,
x^{2(k-h-j)+p} \, y^{2j+|m|-p} \,  z^{2h+l-2k-|m|},
\end{aligned}
\end{equation}
with
\begin{equation}
\mathcal{C}_1(l,|m|,k) = \frac{(-1)^k}{2^l} \binom{l}{k}
\binom{2l-2k}{l} \frac{(l-2k)!}{(l-2k-|m|)!},
\end{equation}
and
\begin{equation}
\mathcal{C}_2(m,p) =
\left\lbrace
\begin{aligned}
& (-1)^m  e^{\imath (m-p)\frac{\pi}{2}}
\quad &\text{if } m \ge 0,\\
& e^{-\imath (|m|-p)\frac{\pi}{2}}
\quad &\text{if } m < 0.
\end{aligned}
\right.
\end{equation}

By using the optimized complex Gaussians representation~(\ref{eq:optimized_cG}) of the radial continuum, and inserting formulae
(\ref{product}) and (\ref{cartesianHarmo}) in (\ref{eq:integ_I_def}), the integral $\mathcal{I}_{l,m}$  is separated  into three one-dimensional integrals,
\begin{equation}
\begin{aligned}
\mathcal{I}_{l,m}
&=
\sqrt{\frac{2l+1}{4 \pi} \, \frac{(l-|m|)!}{(l+|m|)!}}  \, \,
\sum_{k=0}^{[\frac{l-|m|}{2}]}   \mathcal{C}_1(l,|m|,k) \,
\sum_{h=0}^{k} \binom{k}{h} \, \sum_{j=0}^{k-h} \binom{k-h}{j} \\
& \quad \times
\sum_{p=0}^{|m|} \, \binom{|m|}{p} \, \left(\mathcal{C}_2(m,p)\right)^* \,
\sum_{s=1}^N \left[ c_s \right]_{l,k_e}
\, e^{-\frac{\beta_i \left[\alpha_s\right]_l}
	{\beta_i + \left[\alpha_s\right]_l} R^2} \,
\mathcal{I}_x \, \mathcal{I}_y \, \mathcal{I}_z
\text{,}
\end{aligned}
\label{eq:integ_I_xyz}
\end{equation}
where
\begin{equation}
\left\lbrace
\begin{aligned}
&\mathcal{I}_x =
\int_{-\infty}^{\infty} \,
e^{- \left[\gamma_s\right]_l \left( x - \left[ x_{C_s} \right]_l \right)^2 } \,
x^{2(k-h-j)+p} \, \left(x-X\right)^{n_x} \,
dx \text{,} \\
&\mathcal{I}_y =
\int_{-\infty}^{\infty} \,
e^{- \left[\gamma_s\right]_l \left( y - \left[ y_{C_s} \right]_l \right)^2 } \,
y^{2j+|m|-p} \, \left(y-Y\right)^{n_y}  \,
dy \text{,} \\
&\mathcal{I}_z =
\int_{-\infty}^{\infty} \,
e^{- \left[\gamma_s\right]_l \left( z - \left[ z_{C_s} \right]_l \right)^2 } \,
z^{2h+l-2k-|m|+1} \,  \left(z-Z\right)^{n_z} \,
dz \text{.} \\
\end{aligned}
\right.
\end{equation}
Since the integrals $\mathcal{I}_x,\mathcal{I}_y,\mathcal{I}_z$ are
similar, it is sufficient  to evaluate  for example $\mathcal{I}_x$.
To do so, we use the expansion~\cite{taketa1966gaussian}
\begin{equation}
\begin{aligned}
&x^{2(k-h-j)+p} \, \left(x-X\right)^{n_x}  =
\sum_{i_1=0}^{2(k-h-j)+p} \binom{2(k-h-j)+p}{i_1} \\
& \quad \times
\sum_{i_2=0}^{n_x} \binom{n_x}{i_2}
\left[ x_{C_s} \right]_l^{2(k-h-j)+p-i_1}
\left( \left[ x_{C_s} \right]_l - X  \right)^{n_x-i_2}
\left( x - \left[ x_{C_s} \right]_l \right)^{i_1+i_2},
\end{aligned}
\end{equation}
and, after making a variable change substracting $\Re([x_{C_s}]_l)$,
we write
\begin{equation}
\begin{aligned}
\mathcal{I}_x
&=
\sum_{i_1=0}^{2(k-h-j)+p} \binom{2(k-h-j)+p}{i_1}
\sum_{i_2=0}^{n_x} \binom{n_x}{i_2}
\left[ x_{C_s} \right]_l^{2(k-h-j)+p-i_1}   \\
& \quad \times
\left( \left[ x_{C_s} \right]_l - X  \right)^{n_x-i_2}
\int_{-\infty}^{\infty} \,
e^{- \left[\gamma_s\right]_l
	\left( x - \imath \Im\left(\left[ x_{C_s} \right]_l\right) \right)^2 } \,
\left( x - \imath \Im\left(\left[ x_{C_s} \right]_l\right) \right)^{i_1+i_2} \, dx
\text{.}
\end{aligned}
\label{eq:inte_Ix_def}
\end{equation}
We recall here that $\Re\left(\left[\gamma_s\right]_l\right) > 0$.
By a contour integration of the function
$f(z)=z^n \exp \left( -\left[\gamma_s\right]_l z^2 \right)$
over the rectangle
$\left( -L,0 \right) \rightarrow
\left( L,0 \right) \rightarrow
\left( L,\ell \right) \rightarrow
\left( -L,\ell \right) \rightarrow
\left( -L,0 \right)$  in the complex plane, and by taking $L \to \infty$  and
$\ell=-\Im\left(\left[ x_{C_s} \right]_l\right)$, one can easily show that
the imaginary shift does not affect the integral in~(\ref{eq:inte_Ix_def})~\cite{AmmarPhD}.
In other words, the term $\Im\left(\left[ x_{C_s} \right]_l\right)$ can be removed, and
 $\mathcal{I}_x$ is finally given as a double sum of algebraic terms
\begin{equation}
\begin{aligned}
\mathcal{I}_x
&=
\sum_{i_1=0}^{2(k-h-j)+p} \binom{2(k-h-j)+p}{i_1}
\sum_{i_2=0}^{n_x} \binom{n_x}{i_2}
\left[ x_{C_s} \right]_l^{2(k-h-j)+p-i_1}   \\
& \quad \times
\left( \left[ x_{C_s} \right]_l - X  \right)^{n_x-i_2}
\mathcal{Q}_{i_1+i_2}\left(\left[\gamma_s\right]_l\right)
\end{aligned}
\label{eq:inte_Ix}
\end{equation}
with the classical Gaussian integrals
\begin{equation}
\begin{aligned}
\mathcal{Q}_{n}\left(\alpha\right)
&\equiv
\int_{-\infty}^{\infty} \, e^{- \alpha x^2 } \, x^{n} \, dx
=
\left\lbrace
\begin{aligned}
&\frac{\sqrt{\pi} \, n!}{2^n \, (n/2)! \, \alpha^{\left(\frac{n+1}{2} \right)}}
\quad &\text{if } n= \text{ even integer,}\\
& 0
\quad &\text{if } n= \text{ odd integer.}
\end{aligned}
\right.
\end{aligned}
\end{equation}
In a similar manner we can evaluate $\mathcal{I}_y,\mathcal{I}_z$,
and finally obtain	
\begin{equation}
\label{integral with Cartesian}
\begin{aligned}
\mathcal{I}_{l,m}&=
\sqrt{\frac{2l+1}{4 \pi} \, \frac{(l-|m|)!}{(l+|m|)!}}  \, \,
\sum_{k=0}^{[\frac{l-|m|}{2}]}   \mathcal{C}_1(l,|m|,k) \,
\sum_{h=0}^{k} \binom{k}{h} \, \sum_{j=0}^{k-h} \binom{k-h}{j} \\
& \quad \times
\sum_{p=0}^{|m|} \, \binom{|m|}{p} \, \left(\mathcal{C}_2(m,p)\right)^* \,
\sum_{s=1}^N \left[ c_s \right]_{l,k_e}  \,
e^{-\frac{\beta_i \left[\alpha_s\right]_l}{\beta_i + \left[\alpha_s\right]_l} R^2} \\
& \quad \times  \left[
\sum_{i_1=0}^{2(k-h-j)+p} \binom{2(k-h-j)+p}{i_1}  \right. \\
& \quad \times \left.
\sum_{i_2=0}^{n_x} \binom{n_x}{i_2}
\left[ x_{C_s} \right]_l^{2(k-h-j)+p-i_1}
\left( \left[ x_{C_s} \right]_l - X  \right)^{n_x-i_2}
\mathcal{Q}_{i_1+i_2}\left(\left[\gamma_s\right]_l\right)
\vphantom{\sum_{i_1=0}^{2(k-h-j)+p} \binom{2(k-h-j)+p}{i_1}}
\right] \\
& \quad \times \left[
\sum_{j_1=0}^{2j+|m|-p} \binom{2j+|m|-p}{j_1} \right. \\
& \quad  \times \left.
\sum_{j_2=0}^{n_y} \binom{n_y}{j_2}
\left[ y_{C_s} \right]_l^{2j+|m|-p-j_1}
\left( \left[ y_{C_s} \right]_l - Y  \right)^{n_y-j_2}
\mathcal{Q}_{j_1+j_2}\left(\left[\gamma_s\right]_l\right)
\vphantom{\sum_{j_1=0}^{2j+|m|-p} \binom{2j+|m|-p}{j_1}}
\right] \\
& \quad \times \left[
\sum_{k_1=0}^{2h+l-2k-|m|+1} \binom{2h+l-2k-|m|+1}{k_1} \right. \\
& \quad \times \left.
\sum_{k_2=0}^{n_z} \binom{n_z}{k_2}
\left[ z_{C_s} \right]_l^{2h+l-2k-|m|+1-k_1}
\left( \left[ z_{C_s} \right]_l - Z  \right)^{n_z-k_2}
\mathcal{Q}_{k_1+k_2}\left(\left[\gamma_s\right]_l\right)
\vphantom{\sum_{k_1=0}^{2h+l-2k-|m|+1} \binom{2h+l-2k-|m|+1}{k_1}}
\right] ,
\end{aligned}
\end{equation}
where the three double sums within the brackets are independent of each other.


\section{Numerical illustration}

In this section we wish to validate numerically the analytical expressions
derived in the previous sections, whether in spherical or cartesian coordinates.
%
%
%
We consider the transition matrix element
$\mathcal{T}_{\mathbf{k_e}}$~(\ref{eq:transition_element}) by taking
as initial state a 1s-type Gaussian ($n_i=l_i=m_i=0$)
\begin{equation}
\Phi_{i} \left( \mathbf{r} - \mathbf{R} \right)
= \frac{1}{2\sqrt{\pi}}
\exp \left[ -\beta \left( \mathbf{r} - \mathbf{R} \right)^2 \right],
\end{equation}
and as final
continuum state~(\ref{eq:continuum_state}) a pure Coulomb function, with the
choice
$\mathbf{k_e} = k_e \hat{z}$.
This continuum could be replaced by any wave distorted by a realistic potential.

We evaluate numerically the three-dimensional integral
(since the integration over the azimuth angle $\phi$ is trivial,
it is actually a two-dimensional integral that we compute
with \texttt{QUADPACK} \cite{piessens2012quadpack})
whose value, which serves here as reference, is
denoted $\mathcal{T}_{\mathbf{k_e}}^{num}$ hereafter.

To calculate the same matrix element with our method, we first need the complex Gaussian representation of the radial continuum functions (\ref{eq:optimized_cG}).
The continuum state is developed in partial waves;
for each partial term
($l=0,1\dots,L$ with $L=5$), we fit a set of six regular Coulomb functions
$\{ F_{l,k_{e_i}}(r) \}_{i=1,\dots,6}$,
with $N=30$ complex Gaussians, up to a radial distance $r_{\max}=25$ $au$.
The grid of energy is fixed as (in atomic units)
\begin{equation}
k_{e_i} = 0.50 + 0.25 (i-1).
\end{equation}
The non-linear optimization procedure is detailed in \cite{Ammar2020,AmmarPhD}.
We then apply the closed form expressions~(\ref{sphericalME}) with~(\ref{integralnequal0})
and~(\ref{cartesianME}) with~(\ref{integral with Cartesian}) to evaluate the  transition matrix element, whose result is denoted
$\mathcal{T}_{\mathbf{k_e}}^{cG}$.

\begin{figure}
	\center
	\includegraphics[scale=1.10]{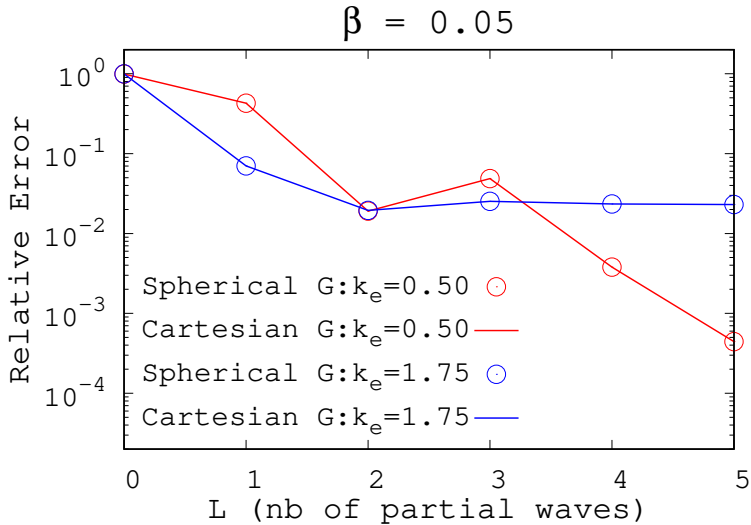}
	\caption{\label{Fig:conv}
	Relative error on $\left| \mathcal{T}_{\mathbf{k_e}} \right|^2$
	in terms of the total number of partial waves of the continuum, for two values of $k_e$, found using either the analytical spherical~(\ref{sphericalME})
or cartesian~(\ref{cartesianME}) expression.
	}
\end{figure}

Calculations were done with an initial state centered on $\mathbf{R} = (0, 0, 1)$ and for several  exponents:
$\beta \in \{0.05, 0.1, 0.5, 1.0\}$.
In figure~\ref{Fig:conv},
we plot the relative error
\begin{equation}
\frac{ \left| \mathcal{T}_{\mathbf{k_e}}^{num} \right|^2
	- \left| \mathcal{T}_{\mathbf{k_e}}^{cG} \right|^2}
	{\left| \mathcal{T}_{\mathbf{k_e}}^{num} \right|^2}
\label{eq:relErr}
\end{equation}
in terms of the number of total partial waves, for $\beta=0.05$ and for two energies ($k_e=0.5$ and $1.75$ a.u.).
Results obtained with the  spherical
and cartesian expressions yield practically the same results.
Even with 6 partial terms, a relative error of $\sim 2\% $
is observed for the higher energy case.
This is related to the relative quality of the representation with 30 complex Gaussians of a fast oscillating continuum,
and the situation will deteriorate with increasing energy.
Note that the increase of the error from $L=2$ to $L=3$ in the $k_e=0.5$ case is not due to the complex Gaussian representation but to a non-monotonic convergence of the partial wave expansion.

\begin{figure}
	\center
	\includegraphics[scale=0.50]{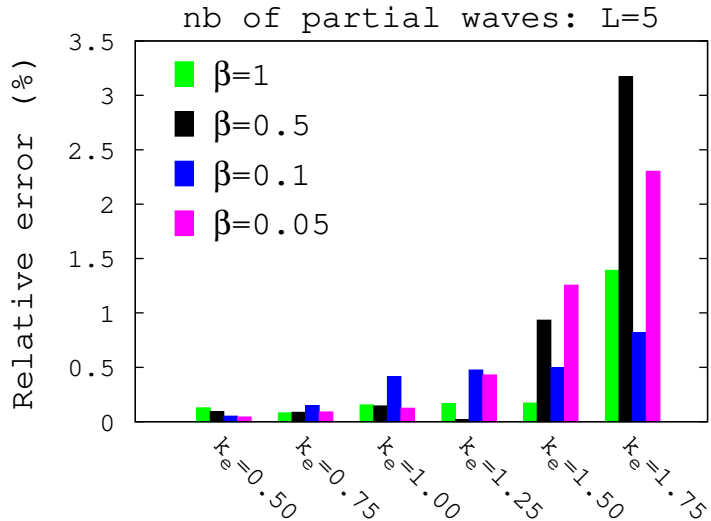}
	\caption{\label{Fig:err_bars}
	Relative error on $\left| \mathcal{T}_{\mathbf{k_e}} \right|^2$
	in terms of the wavenumber $k_e$, for several initial state exponents $\beta$, computed using the analytical cartesian expression~(\ref{cartesianME}).
	}
\end{figure}

The value of the initial state Gaussian decay will regulate the radial extent of the integrand.
Smaller $\beta$ values will require larger domains.
The relative errors~(\ref{eq:relErr}) are plotted in the histogram presented in Figure~\ref{Fig:err_bars}
in terms of the wavenumber $k_e$, and for several initial state exponents
$\beta = 0.05, 0.10, 0.50, 1.00$. The total number of used partial waves
in this calculation is $L=5$ (6 partial waves).
Here the results were obtained by using the cartesian expression~(\ref{cartesianME}) with~(\ref{integral with Cartesian})
(again, practically indistinguishable results were found using the spherical expression~(\ref{sphericalME}) with~(\ref{integralnequal0})).
They are all very good and confirm that bound to continuum transition integrals can be efficiently calculated using complex Gaussian type functions for energy values that are of physical interest.

\section{Conclusions}\label{conclusions}
We have studied three-dimensional multicenter integrals involving a continuum state that appear, for example, in  ionization of molecules in a one-active electron picture.
Through a representation in optimized complex Gaussians for the final continuum, and in real (spherical or cartesian) Gaussians for the initial bound state, we show how matrix elements for photoionization in length gauge can be calculated analytically. The result involves multiple summations. Convergence has been investigated numerically with a test case which showed that the formulation is equally efficient when using a spherical or a cartesian Gaussian description of the bound state.

 This all-Gaussian approach can be formulated similarly for photoionization with the dipole operator in velocity gauge. It can also be adapted to deal with the ionization by impact of a charged particle.

Finally, we intend to go beyond the one-active electron case and tackle
the more interesting, but even more challenging, two-electron integrals.






\end{document}